# High performance Black Phosphorus/Graphitic Carbon Nitride Heterostructure-based Wearable Sensor for Real-time Sweat Glucose Monitoring

Ecem Ezgi Özkahraman[a,b], Zafer Eroğlu[b], Vladimir Efremov[b], Arooba Maryyam[c], Taher Abbasiasl[d], Ritu Das[e], Hadi Mirzajani[e], Berna Akgenc Hanedar[g,h], Levent Beker[d,e]*, Onder Metin[a,b,f]*

[a]Department of Materials Science and Engineering, College of Engineering, Koç University, 34450 Istanbul, Turkiye

[b]Department of Chemistry, College of Sciences, Koç University, 34450 Istanbul, Turkiye

[c]Department of Electrical and Electronics Engineering, College of Engineering, Koç University, 34450 Istanbul, Turkiye

[d]Department of Biomedical Sciences and Engineering, College of Engineering, Koç University, 34450 Istanbul, Turkiye

[e]Department of Mechanical Engineering, College of Engineering, Koç University, 34450 Istanbul, Turkiye

[f]Koç University Surface Science and Technology Center (KUYTAM), 34450 İstanbul, Türkiye

[g]Department of Physics, College of Sciences, Koç University, 34450 Istanbul, Turkiye

[h]Department of Physics, College of Sciences, Kirklareli University, 39100 Kirklareli, Turkiye

*To whom should be corresponded: Prof. O. Metin, e-mail: ometin@ku.edu.tr, Prof. L. Beker, e-mail: lbeker@ku.edu.tr

**Abstract**

Wearable, non-invasive glucose sensors capable of accurate and continuous monitoring are crucial for managing metabolic conditions yet achieving high sensitivity and stability in these devices remains challenging. In this work, we present a Black Phosphorus/Graphitic Carbon Nitride (BP/g-CN) heterostructure engineered to leverage phosphorus-nitrogen interactions for enhanced electrochemical glucose oxidation activity. Compared to pristine gCN, the BP-gCN heterostructure demonstrates a significantly improved electrochemical surface area (ECSA) and nearly two-fold reduction in charge transfer resistance ($R_{ct}$), achieving remarkable glucose sensitivity of 1.1 µAmM$^{-1}$cm$^{-2}$ at physiological pH. Density functional theory (DFT) calculations revealed stronger glucose adsorption and higher charge transfer on the BP-gCN heterostructure compared to pristine gCN surface. These theoretical insights complement the experimental findings, highlighting the superior electrocatalytic performance of the heterostructure and the role of oxidized BP surface. Furthermore, the BP-gCN sensor is integrated into a wearable device platform with microfluidic layers and a Near Field Communication (NFC) chip, forming a conformal skin patch that enables real-time sweat glucose monitoring. This demonstration of a high-performance, non-enzymatic wearable glucose sensor based on a heterostructure design underscores the potential of the device for seamless health management and paves the way for next generation biosensing platforms aimed at improving personalized and continuous health monitoring.

# 1. Introduction

In recent years, wearable biosensors have emerged as transformative platforms for personalized healthcare, enabling continuous, noninvasive physiological biomarker monitoring and delivering real-time feedback to users outside traditional clinical settings.[1-3] Rapid progress in flexible materials, scalable manufacturing processes, wireless communication protocols, and miniaturized electronics have accelerated the development and deployment of advanced wearable devices.[4,5] Despite this progress, current commercial wearable sensors predominantly monitor basic parameters like heart rate and body temperature, thus failing to capture insightful molecular-level information crucial for comprehensive health assessment. Recent research has focused on integrating chemical sensors into wearable devices capable of analyzing biological fluids such as interstitial fluid (ISF), tears, saliva, and sweat, providing detailed information on various key biomarkers.[6-10] For instance, diabetes mellitus, a chronic disease characterized by irregular glucose levels in the blood, is typically monitored using glucometers.[11] These devices require users to prick their fingers multiple times a day to measure blood glucose levels, which does not provide continuous monitoring. In this regard, chemical biosensors integrated into wearable devices, such as skin patches, tattoos, and contact lenses have been proposed for continuous and painless monitoring of glucose levels.[12-14] Among these strategies, sweat-based glucose sensors are particularly promising, as they leverage a naturally secreted, easily accessible fluid, enabling truly noninvasive and continuous glucose monitoring with minimal user discomfort.[15]

To date, a wide range of measurement techniques including colorimetric, fluorescent, photoelectrochemical and electrochemical methods have been explored for detecting glucose in sweat.[16–19] Among these methods, electrochemical sensing is particularly attractive for wearable applications due to its rapid response, low power requirements, ease of miniaturization, high sensitivity, selectivity, and straightforward fabrication processes.[20] However, majority of the reported wearable electrochemical glucose sensors require the use of an enzyme, glucose oxidase (GOx) or glucose dehydrogenase (GDH), which converts the glucose oxidation into an electrical signal.[21,22] While enzymatic sensors offer excellent selectivity, their stability is often compromised by environmental fluctuations in temperature and pH, ultimately reducing their operational lifespan and reliability in real-world, on-body conditions.[23] Moreover, the high cost of enzymes significantly hinders the scalability and long-term affordability of enzymatic sensors, posing a serious economic barrier to widespread adoption in wearable healthcare systems. To address these challenges, there is a growing emphasis on the development of stable, robust, and cost-effective novel materials capable of non-enzymatic glucose oxidation.

Recently, a variety of materials, including metals,[24,25] metal oxides,[26,27] alloys[28,29] and carbon nanocomposites[10,30] have been investigated for non-enzymatic glucose oxidation. While these materials have demonstrated promising performance in the glucose detection, they suffer from significant challenges, such as poor selectivity/sensitivity, surface poisoning, and high costs. Additionally, the use of metal-based materials for the wearable sensors have several drawbacks, including non-biocompatibility and the need for high pH conditions as the metal surface needs hydroxide ions for the glucose oxidation reaction.[31] To address these challenges, researchers have focused on developing methods that generate a localized alkaline environment, facilitating the use of metallic materials as non-enzymatic glucose sensor.[32,33] However, these approaches require high voltage for sensor activation, which poses drawbacks in terms of energy efficiency and necessitates additional pretreatment steps. Consequently, metal-free carbon-based materials have recently gained attention for their superior electron transfer kinetics and resistance to surface poisoning.[34,35] Nevertheless, pure carbon-based materials such as graphene are typically inactive toward the oxidation of alcoholic groups, limiting their effectiveness for electrochemical glucose sensing.[36] Therefore, exploring novel materials such as graphitic carbon nitride (gCN) and its heterostructures with other non-metallic semiconductors is crucial for wearable non-enzymatic glucose sensing, as these materials show promise for advancing the field.

gCN, a polymeric semiconductor composed of C and N atoms, is promising for wearable sensors owing to its 2D structure, large surface area, biocompatibility, and chemical stability.[37,38] gCN is known for its intrinsic peroxidase-like activity[39] and its applications for electrochemical glucose sensing have been previously investigated.[40-42] Despite the promising potential of gCN for glucose oxidation, significant challenges remain including high electron-hole recombination rates and low electrocatalytic activity, which we aim to address through innovative heterostructure design with complementary non-metallic materials. On the other hand, black phosphorus (BP), a mono-elemental 2D semiconducting material composed of earth-abundant phosphorus atoms, features a tunable bandgap ranging from 0.3 to 2.1 eV, high charge carrier mobility, and excellent biocompatibility, making it particularly suitable for sensor applications.[43] The electrocatalytic activity of BP and its composites has been widely explored, with focus on the enhanced electrocatalytic performance of oxidized BP (oBP). The superior stability of oBP in glucose dehydrogenation further supports its potential in sensor applications.[44] Additionally, the high surface area of BP nanosheets contributes to increased oxidation current, further enhancing its effectiveness in electrochemical sensing.[45] The oxidation process generates P=O redox sites, enhancing electron transport and improving overall sensor performance, especially when combined with materials such as reduced graphene oxide (rGO).[46] Recent research emphasizes the promise of BP-gCN heterostructures, where P-N coordination significantly enhances charge transfer by trapping the electron in defects at their interfaces[47] suggesting them as strong candidates for electrochemical sensing. Despite these promising attributes and recent advancements, the full potential of BP-gCN heterostructure for non-enzymatic glucose oxidation in wearable configuration remains unexplored. This gap presents a crucial opportunity to develop a robust, sensitive and stable electrochemical sensor for glucose detection, which could significantly enhance continuous glucose monitoring capabilities in wearable devices.

In this study, we present a novel approach that integrates rational material design with wearable sensor technologies to develop a flexible, wearable non-enzymatic electrochemical glucose sensor based on the BP-gCN heterostructures, capable of operating at physiological pH for sweat glucose detection with a sensitivity of 1.1 µA mM$^{-1}$ cm$^{-2}$. Initially, BP-gCN heterostructures with varying BP-gCN ratios were prepared to determine the optimal composition for enhancing sensor activity, followed by characterizing them with advanced techniques. The structural characterization of the BP-gCN heterostructures revealed that the constructed P-N interactions in BP-gCN heterostructure improved the glucose oxidation by increasing electrochemical active surface area (ECSA) and reducing charge transfer resistance ($R_{ct}$) almost two-fold, compared to pristine gCN, prompting us to investigate this superior glucose sensing results through density functional theory (DFT) calculations. The glucose adsorption on gCN and BP-gCN was compared, revealing the stronger adsorption and higher charge transfer on the latter surface, which suggests the improved catalytic efficiency for glucose oxidation. Notably, the surface oxides on BP provide additional redox-active sites, which further enhance the glucose oxidation process and contribute to the superior electrochemical performance of the heterostructure. To demonstrate in-vivo applicability, the synthesized heterostructures were drop-cast onto screen-printed carbon electrodes (SPCE) and integrated with electronics and microfluidic layers to develop a wearable skin patch. On-body electrochemical evaluations were conducted wirelessly, with real-time monitoring performed via a smartphone using a commercial NFC chip.

## 2. Results and Discussion

### 2.1. Synthesis and Characterization of BP-gCN Heterostructures

To construct a non-enzymatic wearable glucose sensor, we designed a 2D material based heterostructures composed of BP and gCN as the electrochemical sensor layer, integrated between a microfluidic sweat collection platform and a wireless communication module equipped with NFC technology, as illustrated in **Figure 1a**. First, gCN was synthesized according to the well-established thermal polycondensation method while BP was fabricated via mineralizer assisted chemical vapor transport strategy. Following, ultrasound-assisted liquid phase exfoliation approach was utilized to prepare the targeted BP-gCN heterostructures. To investigate the superiority of the heterostructures, the loading ratio of BP on gCN was systematically varied to evaluate its synergistic effect as an electrocatalyst.

To gain comprehensive insight into the crystallinity and phase composition of the as-prepared materials, powder-XRD analysis was employed. As shown in **Figure 1c**, two characteristic main peaks of gCN are observed at $2\theta = 27.3°$ (diffraction from the (002) plane) for the stacking of conjugated tri-s-triazine rings and at $2\theta = 13.2°$ (diffraction from the (100) plane) for in-plane interactions.[38] Moreover, the XRD pattern of BP confirms the orthorhombic crystal structure, with the diffraction peaks at $2\theta = 16.9, 34.2,$ and $52.5$, corresponding to the (020), (040), and (060) planes, respectively (JCPDS card no. 73-1358).[48] After investigating the crystallinity of pristine materials, XRD patterns of binary heterostructures were collected and compared with them. **Figure 1c** shows an increasing trend in the BP peak around $2\theta = 34.5°$ (diffraction from the (040) plane) and the peak at $2\theta = 17°$ (diffraction from the (020) plane)[49] in parallel with the increment of loading ratio, as well as the slight shift of the gCN (100) facet peak to lower angles, indicates a decrease in crystallite size and an increased in lattice strain. These results confirm the successful construction of targeted BP-gCN binary heterostructure

In the Raman spectrum given in **Figure 1d**, three characteristic Raman shifts of BP were observed. These Raman shifts correspond to the out-of-plane phonon mode $A_g^1$ at 360.9 cm$^{-1}$, the in-plane phonon modes $B_{2g}$ and $A_g^2$ at 438.6 cm$^{-1}$ and 465.5 cm$^{-1}$, respectively.[44] More importantly, three peaks around 800-900 cm$^{-1}$ are assigned to the Raman active modes of phosphorus oxides, confirming the oxidized form of BP .[50] The surface oxides on BP contribute significantly to glucose oxidation by creating additional redox-active sites and facilitating an efficient electron transfer between glucose molecules and the catalyst surface, leading to an improved electrochemical response. This increase in electrochemical performance emphasizes the critical role of surface oxides in optimizing glucose sensing applications.

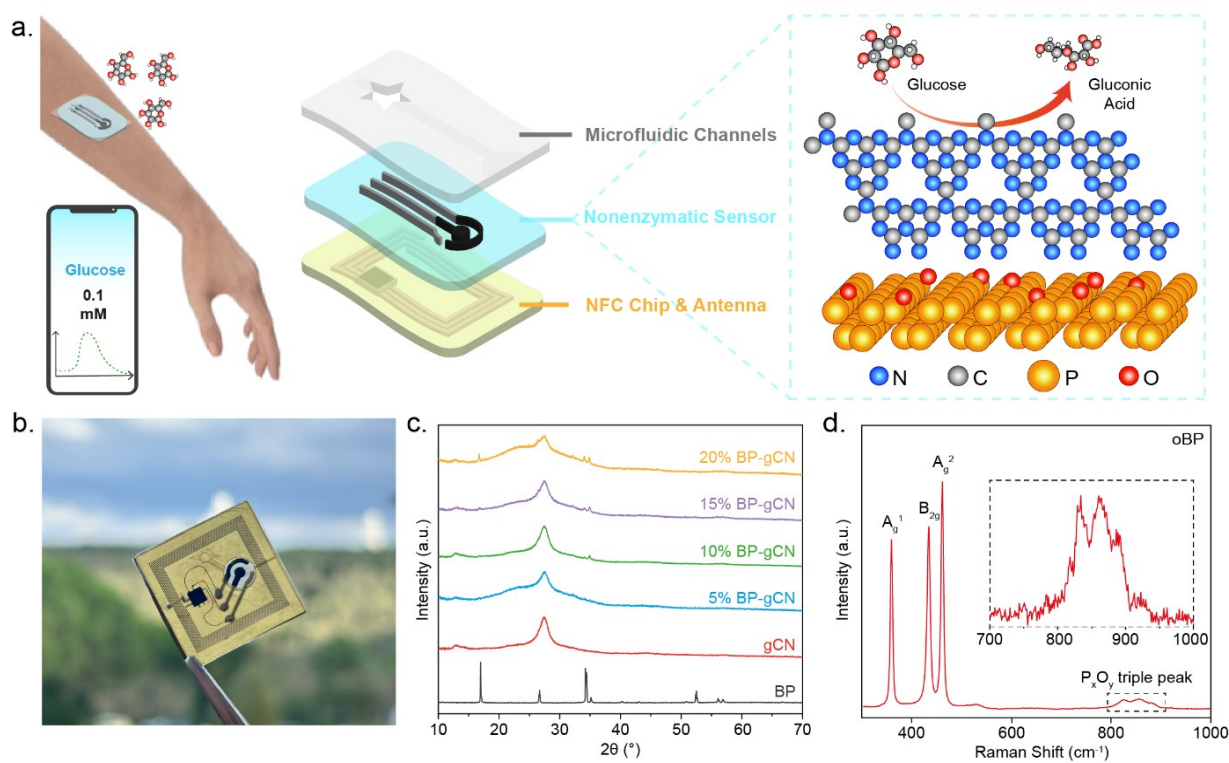

**Figure 1. Schematic illustration of the nonenzymatic patch and structural characterizations. a)** Designed wearable non-enzymatic sweat patch with layers including microfluidic channels for sweat collection (left), non-enzymatic sensor array and wireless communication layer by NFC (middle), schematic representation of BP-gCN showing the oxidation of glucose to gluconic acid on the surface (right). **b)** Photograph of the fabricated device. **c)** XRD patterns of gCN and 5, 10, 15, 20 wt% BP-gCN **d)** Raman spectra of oBP.

The UV-VIS absorption of the BP-gCN heterostructures was measured, and the bandgaps were determined by using Tauc plots, as shown in **Figure S1** (**Supporting Information**). The analysis reveals that the addition of BP to gCN results in a narrowed bandgap, which indicates improved charge separation. This reduction in bandgap contributes to the observed decrease in $R_{ct}$, as a narrower bandgap facilitates more efficient electron transfer pathways. Improved charge separation and reduced recombination rates are essential for maintaining a high and stable current response in electrochemical sensing. These characteristics suggest that the oxidized BP-gCN heterostructure can effectively facilitate electron transfer between the sensing surface and the target analyte, thereby enhancing detection sensitivity by enabling faster and more efficient charge transfer.

Furthermore, XPS analysis was performed to confirm the oxide formation on the BP surface and the surface chemistry of the BP-gCN heterostructure (**Figure 2d,e**). XPS survey spectrum of the BP-gCN (**Figure S2**, **Supporting Information**) confirms the presence of all expected elements (C, N, P, O) and the absence of any impurities in the heterostructure. In the high-resolution XPS spectrum of C 1s region, the peak at 284.5 eV was attributed to the standard reference carbon, while the peak at 286.2 eV corresponds to carbon atoms bonded to nitrogen (sp$^3$-hybridized), bridging the heptazine rings in gCN. The dominant peak at 287.9 eV is associated with the C=N bonding in the triazine ring of gCN. A shift of 0.13 eV towards higher binding energies in the C=N peak for BP-gCN suggests an interaction between BP and gCN, likely via electron transfer from gCN to BP.[51] In the high-resolution XPS spectra of N 1s region, the peak corresponding to the C=N bonds in the s-triazine aromatic rings of the heterostructure was observed at 398.48 eV compared to 398.42 eV for pristine gCN, indicating a shift of 0.06 eV to higher binding energies. This shift suggests electron redistribution and a strong electronic interaction between BP and gCN. Additionally, N-(C)$_3$ peak was detected at 399.9 eV for BP-gCN and 400.1 eV for gCN, showing a notable 0.2 eV shift to lower binding energies, further confirming the interaction between BP and gCN. Meanwhile, the C-N peak at 401.1 eV appeared without any alteration, indicating that these groups remain largely unaffected in the heterostructure.[52] In the high-resolution XPS spectra of P 2p region, two distinct peaks appeared at 129.6 eV and 130.5 eV, corresponding to the P-P bonds of $2p_{3/2}$ and $2p_{1/2}$, respectively. Additionally, in BP-gCN heterostructure, a broad oxidation peak at 134.1 eV is readily assigned to P-O bond of native surface oxide $P_xO_y$ species, while a peak at 133.1 eV is linked to P-N interaction.[47] Based on the XPS peak area, $P_xO_y$ species account for approximately 20% coverage of the BP surface. Compared to pristine BP, a shift to lower binding energies is spotted in the BP-gCN heterostructure, consistent with results seen in the high-resolution XPS spectra of C 1s and N 1s regions. The binding energy shifts observed in the C 1s, N 1s, and P 2p regions, along with the formation of P-N interaction and charge transfer from gCN to BP, indicate that the heterostructure is primarily governed by electronic interactions and physisorption. These shifts reflect electron redistribution and interfacial interactions, involving charge transfer and localized bonding, which contribute to the stability of the heterostructure. Such interactions make the BP-gCN heterostructures highly suitable for applications in electrochemical sensors. Moreover, the observed P-N interactions and formation of oBP enhance the structural integrity, further supporting its potential as an electrochemical sensing platform.[53]

After characterizing the structural properties of all materials, we proceeded to examine their morphological and topological features using SEM, TEM, and AFM analyses. SEM images of gCN revealed a 2D lamellar structure with irregularly curved folds at the edges, while BP exhibited a flake-like morphology with a smooth surface, as shown in **Figure S3 (Supporting Information)**. In the BP-gCN heterostructure, both materials retained their distinct structural features without any topological changes, indicating the formation of a strong interaction between them, as depicted in **Figure 2a**. Consistent with the SEM observations, TEM analysis revealed a layered, mesoporous polymeric structure for gCN, while BP displayed a flat, nanoleaf-like structure (**Figure S4**, **Supporting Information**). The TEM image of BP-gCN (**Figure 2b**) further confirmed the interaction between the two components at nanoscale, with both morphologies remaining intact. Additionally, further TEM images in **Figure S5** (**Supporting Information**) support this nanoscale integration. According to AFM analysis, the thickness of the as-prepared materials was measured to be 19 nm for gCN and 50 nm for BP, as shown in **Figure S6** (**Supporting Information**). The lateral size of the BP-gCN heterostructure (**Figure 2c**) decreased to 4 nm at the nanoscale compared to gCN, while it increased compared to BP.

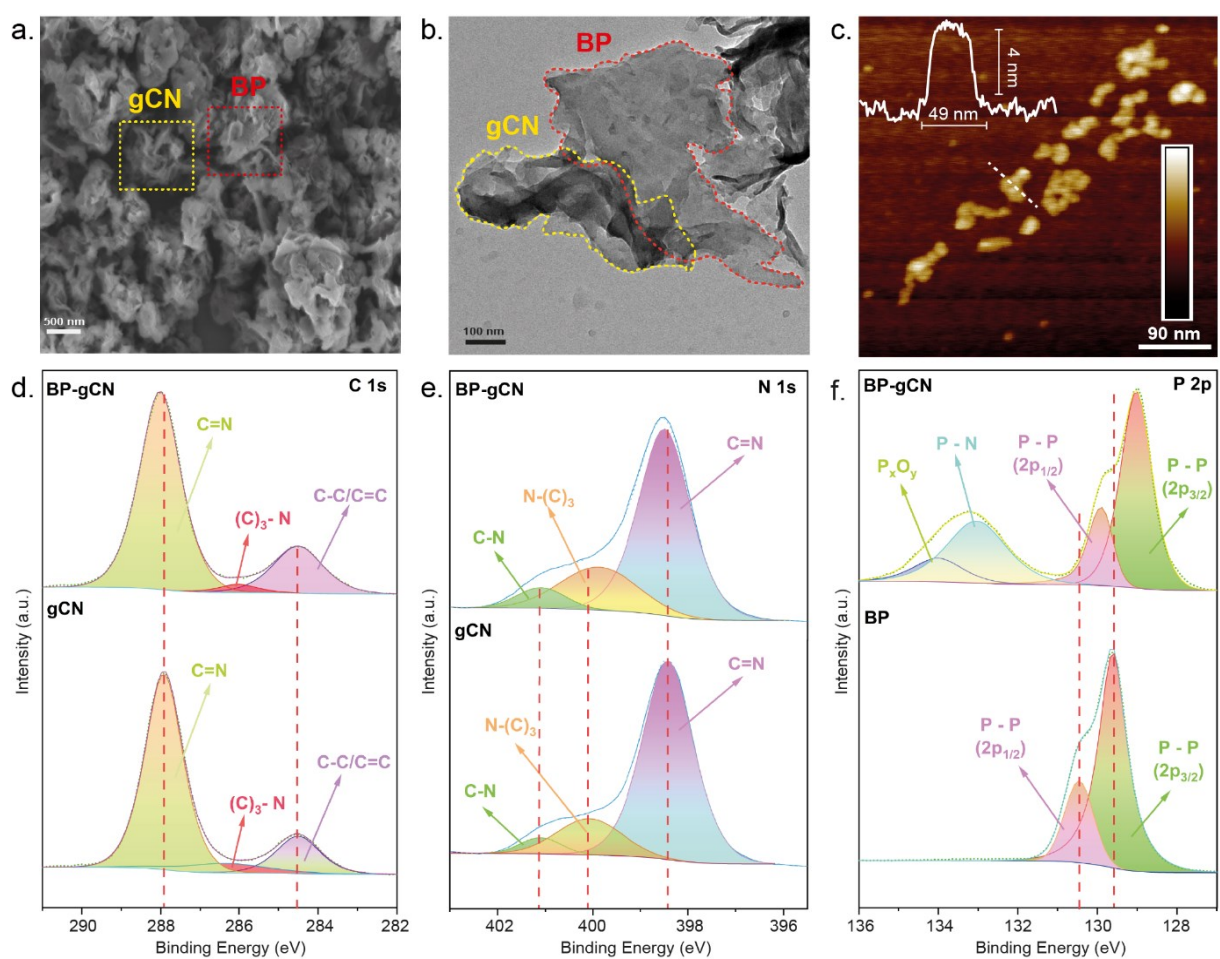

**Figure 2. Characterization of pristine BP, pristine gCN and BP-gCN heterostructure.**
**a)** SEM image **b)** TEM image and **c)** AFM image with the inset of height profile of the BP-gCN heterostructures. XPS spectrum of **d)** C1s, **e)** N 1s, **f)** P 2p for pristine gCN, pristine BP (bottom) and BP-gCN heterostructure (top).

## 2.2. Performance and Characterization of the Electrochemical Sensor

Following material characterization, the intrinsic electrochemical properties of various BP-gCN compositions (5, 10, 15, and 20 wt% BP) were evaluated by conducting Cyclic Voltammetry (CV) measurements in a standard redox probe solution of $K_3[Fe(CN)_6]$ (in 0.1 M KCl) over a potential range of -0.4 V to -0.2 V at a scan rate of 0.02 V/s, providing a baseline for assessing electron transfer efficiency. As illustrated in the CV graphs (**Figure 3a**) and the corresponding peak current bar graph (**Figure 3b**), 10 wt% BP-gCN heterostructure showed the optimal charge transfer and active site availability, resulting in the highest current density and best peak resolution among the tested compositions. This result indicates that the 10 wt% BP-gCN composition optimizes electron transfer, which is essential for achieving high electrochemical sensor performance. The improved charge transfer kinetics arise from the synergistic effect of newly formed redox-active sites on oBP, as confirmed by Raman and XPS analyses, combined with the large surface area of gCN, collectively enabling more efficient electron transport and abundant active sites for glucose oxidation.

To emphasize the critical role of incorporating BP to gCN, we compared the Electrochemical Impedance Spectroscopy (EIS) and ECSA. The Nyquist plots obtained from EIS measurements were used to compare the charge transfer resistance of gCN (red) and BP-gCN (black) (**Figure 3c**). The inset illustrates the equivalent circuit model used to fit the EIS Nyquist plot, which includes several key elements: the resistance of the electrolyte solution ($R_s$), the Warburg element associated with the diffusion of the redox probe, the constant phase element (CPE), and $R_{ct}$. The semicircle of the BP-gCN is smaller than that of pristine gCN, indicating almost two times lower $R_{ct}$ upon BP incorporation. This improvement can be attributed to the synergistic effects of BP and gCN, which facilitates better electron transport and conductivity in the heterostructure. The overall lower impedance of BP-gCN suggests enhanced electrochemical performance, as lower impedance correlates with higher sensor sensitivity. Additionally, the ECSA was calculated using the Randles-Sevcik equation based on CV measurements at different scan rates for gCN (**Figure 3d,e**) and BP-gCN (**Figure 3f,g**). By substituting the relevant parameters into the equation, the results demonstrated that BP-gCN ECSA is 1.6 times greater than that of pristine gCN, offering a greater number of active sites for glucose adsorption and subsequent oxidation. The comparative analysis of EIS and ECSA highlights the significant enhancements achieved by the formation of heterostructure between BP and gCN. The formation of heterostructure not only increased active sites but also improved electron transfer kinetics, collectively contributing to the superior electrochemical performance observed in this study.

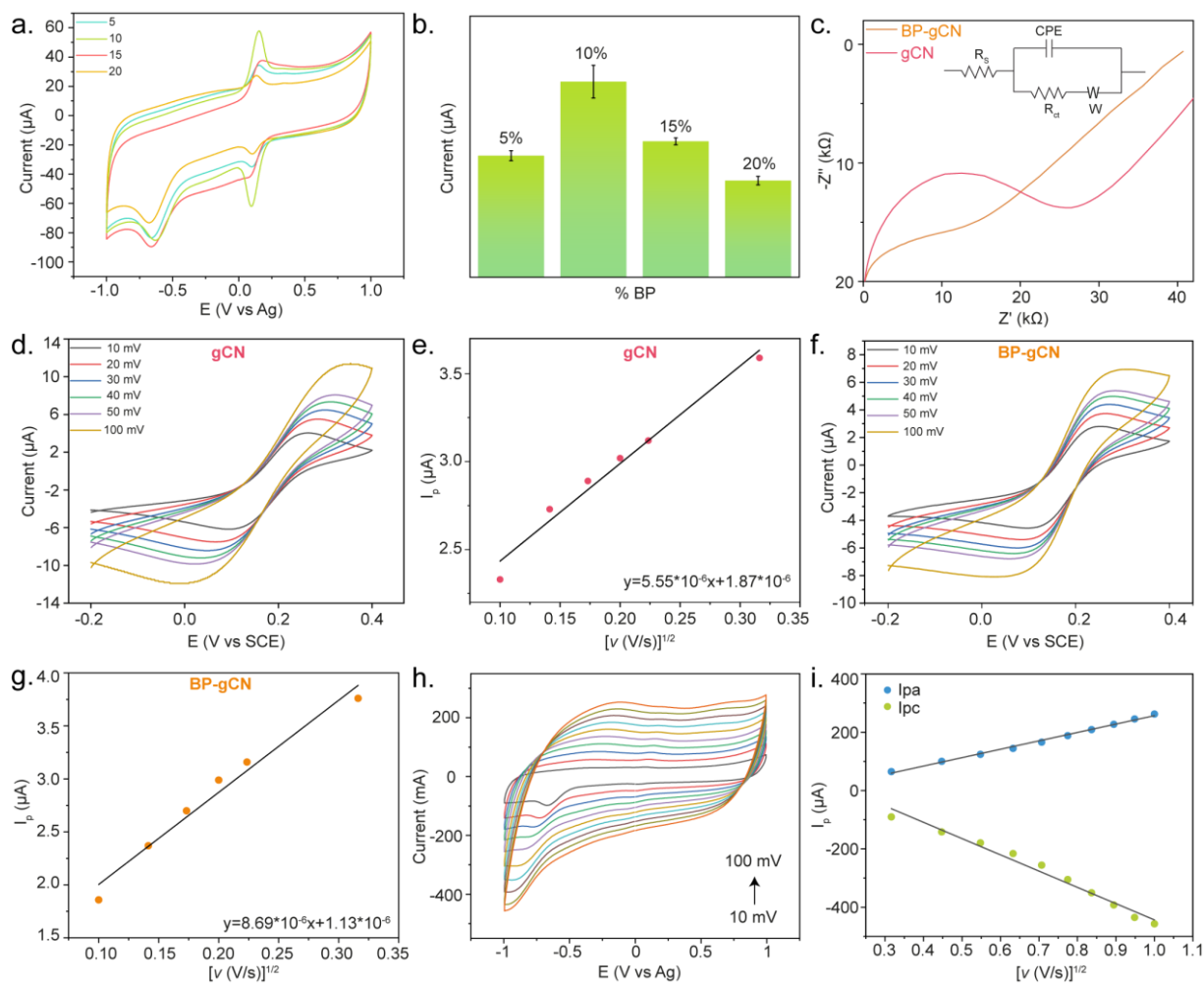

**Figure 3. The electrochemical characterization of pristine gCN and BP-gCN heterostructure. a)** CV comparison of (5-20 %wt) BP-gCN and **b)** corresponding bar graph for electrocatalytic activity at $Ip_a$ **c)** Nyquist plot of gCN and BP-gCN (inset equivalent circuit model used for the fitting). CV graphs at various scan rates for **d)** gCN and **e)** BP-gCN in $K_3[Fe(CN)_6]$. **f)** CV graphs at various scan rates for BP-gCN in 1 mM glucose. Corresponding linear fittings for Randles-Sevcik equations; **g)** gCN, **h)** BP-gCN, **i)** BP-gCN-1 mM glucose.

To further gain a mechanistic understanding of how the glucose molecules interact with the BP-gCN electrocatalyst surface, CV measurements were conducted at various scan rates in 1 mM glucose (in PBS) solution (**Figure 3h,i**). The increase in both maximum and minimum peak currents with increasing scan rates suggests that the electrochemical process is diffusion-controlled, as supported by the linearity of the curve extracted from scan rate effect in accordance with the Randles-Sevcik equation. Complementing these findings, lower frequency region of EIS plot shows the presence of Warburg element, which signifies the effect of diffusion limitation on overall electrochemical reaction kinetics. The Warburg element of BP-gCN has a reduced slope compared to that of gCN, indicating reduced diffusion limitations, improved mass transport, and consequently a faster electrochemical response.

After determining the optimal composition, further electrochemical characterizations were conducted with the 10 wt% BP-gCN modified electrode in PBS (pH=7.0) to evaluate the sensitivity, selectivity, stability, and concentration dependence of the sensor towards glucose oxidation under physiological conditions ex-situ. Firstly, the electrocatalytic activity of modified and un-modified electrodes were compared in 5 mM $K_3[Fe(CN)_6]$ (0.1 M KCl), showing increased activity for modified electrode (**Figure 4a**). For the glucose sensing experiment, CV was employed to identify possible oxidation potential of glucose (1-10 mM in PBS) in the anodic scan. According to **Figure 4b**, oxidation of glucose starts after 0 V and peak separation at the lowest potential is observed at 40 mV. Following this, **Figure 4c** illustrates the representative sensing mechanism, detailing the oxidation of glucose to gluconic acid. The oxidation products were confirmed through quadrupole time-of-flight mass spectrometry (qTOF-MS), elucidating the electrochemical pathways involved. According to qTOF-MS spectra (**Figure S7, Supporting Information**), the main product of glucose oxidation is observed as gluconic acid at 219.04 m/z, indicated by the highest peak intensity.

Differential pulse voltammetry (DPV), a sensitive technique ideal for detecting low-concentration analytes, was performed between -0.2 V and +0.4 V on glucose solutions ranging from 0 to 1 mM in 0.1nM increments -levels relevant to physiological sweat glucose- enabling precise quantification of glucose oxidation currents as shown in **Figure 4d**.[54] Additionally, a calibration curve obtained from DPV graphs (**Figure 4e**) at 40 mV was plotted with n=3. The slope of the calibration curve was used as the sensitivity output as suggested by IUPAC[55] and the calculated sensitivity of the proposed sensor was obtained as 1.1 µA mM$^{-1}$ cm$^{-2}$.

To gain a deeper understanding of the sensor's performance characteristics, we assessed the stability and selectivity of the fabricated sensors. The stability of the 10% BP-gCN-based sensor was evaluated through CV over 50 cycles, demonstrating a peak current retention of 84% as illustrated in **Figure S8 (Supporting Information)**. The long-term stability of the BP-gCN is attributed to the surface oxide layer on BP, which forms P-O bonds through partial oxidation while protecting the underlying BP.[56,57] Furthermore, the formation of P-O bonds enhances the surface hydrophilicity, making it suitable for sensor applications.

Given the complex nature of human sweat, which contains various common electroactive interferents, it is great of importance that a sensor used in a wearable device can selectively and accurately sense glucose. Therefore, we also examined the selectivity of the sweat analyte sensor. **Figure 4f** shows that adding various interfering species (ascorbic acid, urea, sucrose, glycine, NH$_3$) to a 1 mM glucose solution resulted in negligible changes in the sensor response, demonstrating excellent anti-interference capability. Typically, non-enzymatic sensors require high working potentials, where more electroactive interferents such as ascorbic acid also oxidize; however, this issue was addressed in this work by employing a low working potential for the sensor. Additionally, the selective behavior of the BP-gCN can be attributed to the heterostructure's negative surface charge[58,59], which promotes selective adsorption of glucose while repelling negatively charged species present in the sweat.

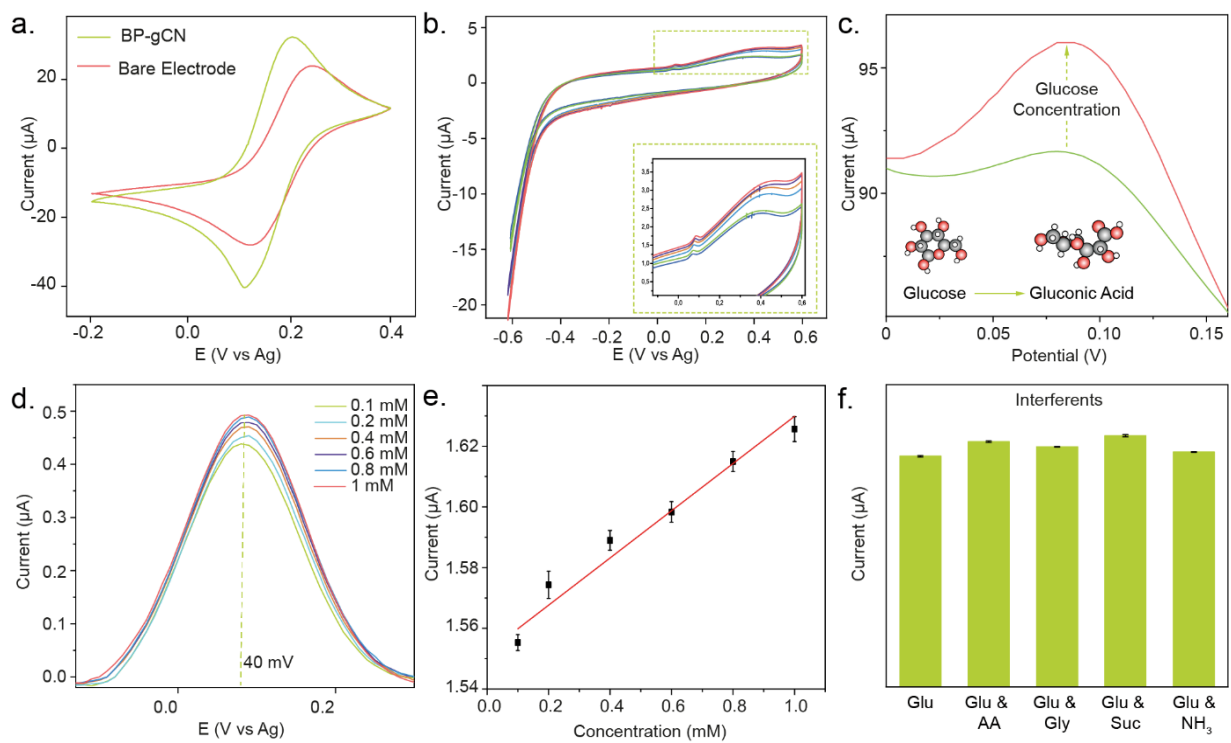

**Figure 4. The electrochemical glucose sensing performance of BP-gCN. a)** CV of 10% BP-gCN modified and bare SPCE in $K_3[Fe(CN)_6]$ between -0.2 V and +0.4V. **b)** Response of sensor to glucose by CV in 0.1 X PBS solution (inset: anodic scan zoomed) **c)** Representation of glucose sensing mechanism of the proposed sensor. **d)** Response of sensor to glucose by DPV. **e)** Corresponding calibration curve at 40 mV. **f)** Selectivity of the sensor with main interferents (ascorbic acid, glycine, sucrose, $NH_3$) showing a slight change in the sensor's response compared to glucose.

## 2.3. Wearable Tag Fabrication and On-Body Evaluation

The fabricated non-enzymatic wearable patch was evaluated for system-level functionality and in vivo compatibility. Before on-body testing, initial benchtop experiments were conducted in an artificial sweat solution for various glucose concentrations. **Figure 5a** demonstrates the correlation between the calibration curves obtained from the potentiostat and the NFC development kit, confirming the ability of the BP-gCN sensor to be integrated into wearable devices for glucose detection in sweat. Building on our earlier report, where we detailed the fabrication process for wearable enzymatic glucose sensors,[60] a similar approach was applied in this work, allowing for a seamless transition to non-enzymatic sensing applications. Briefly, the wearable device is composed of several key components: an NFC chip that handles both sensor reading and data processing, an antenna that provides wireless power and facilitates data communication with a smartphone and microfluidic channels, which transport sweat from the skin to the glucose sensor. The in-vivo physical exercise test, depicted in **Figure 5b**, shows the successful application of the metal-free, non-enzymatic wearable sensor for real-time glucose monitoring in sweat during treadmill running. Initial readings were obtained while the subject was running before consuming any food or drink. After these readings, the subject consumed orange juice at the 30[th] minute and resumed running on the treadmill. New data were then collected to analyze the changes in the glucose profile in sweat resulting from the orange juice intake. The increase in sweat glucose level was observed 1h after consumption, consistent with the typical delay in the response between blood and sweat glucose concentrations. These findings demonstrate the device's applicability in in-vivo settings. With the integration of NFC technology, it enables continuous, real-time glucose monitoring, offering a promising solution for nonintrusive and wearable glucose analysis under physiological conditions.

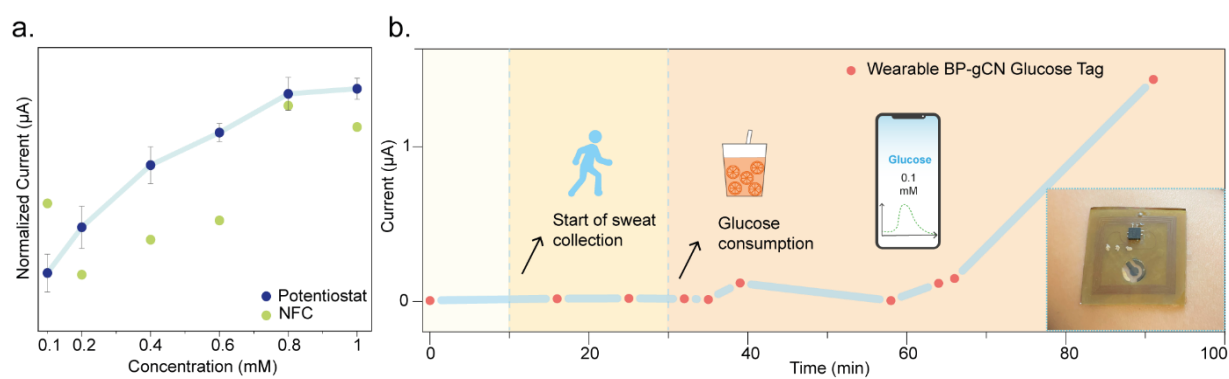

**Figure 5. Integration of BP-gCN to wearable NFC based glucose monitoring patch. a)** Correlation between bench-top potentiostat and NFC development kit for glucose in artificial sweat. **b)** Results of on-body experiment, showing a dramatic increase after orange juice intake at the 90[th] minute (Inset image: Non-enzymatic wearable glucose tag worn on subject's arm).

## 2.4. Computational Study

To investigate and compare the glucose sensing performances of gCN and BP-gCN, DFT calculations were conducted. Firstly, the planar structures of pristine gCN and BP surfaces were optimized and a model of the BP-gCN heterostructure was constructed as illustrated in **Figure S9 (Supporting Information)**. It can be seen that both pristine gCN and BP monolayers become wrinkled when a heterojunction is formed between them. This wrinkling occurs as a result of the strong interactions due to P-N bond formation, which are not present in their pristine forms. This structural transformation can enhance the surface properties of the heterojunction, potentially promoting stronger glucose adsorption, which accounts for the electrocatalytic activity for glucose oxidation.[61,62] Therefore, we examined the glucose adsorption on the planar gCN surface and found that this interaction also induced a structural transformation of the gCN layer from planar to wrinkled (**Figure S10a, Supporting Information**). As illustrated in **Figure S10b (Supporting Information)**, this wrinkled structure persisted even after the glucose molecule was removed. Interestingly, a previous study also reported an irreversible transformation of the gCN monolayer from a planar to a wrinkled structure after the $C_{60}$ adsorption.[63] It was also determined that the wrinkled gCN structure is more stable than the planar structure, with its energy being 4.55 eV lower. Therefore, we proceeded with our DFT calculations using wrinkled gCN and subsequently calculated the glucose adsorption energy ($\Delta E$) on this structure, obtaining a value of -1.04 eV.

As previously discussed in the experimental section, XPS and Raman analysis of the BP-gCN heterostructures revealed the formation of native surface oxides. Based on these findings, we investigated the role of O species on the BP surface. To represent the oxidized BP-gCN catalyst, we constructed the surface-oxidized BP-gCN-$O_{14}$ structure, following the approach of previous studies.[53,64] Glucose adsorption was compared between BP-gCN-$O_1$, a heterostructure with a single oxygen atom, and BP-gCN-$O_{14}$, which features multiple surface oxygen atoms. As shown in **Figure S11 (Supporting Information)**, BP-gCN-$O_1$ does not significantly stabilize the glucose molecule compared to BP-gCN-$O_{14}$, highlighting the role of increased surface oxygen coverage in enhancing glucose adsorption. Notably, our calculations show that glucose exhibits significantly stronger adsorption on BP-gCN-$O_{14}$ compared to wrinkled gCN, with $\Delta E$ of -1.42 eV and -1.04 eV, respectively, as shown in **Figure 6a,b.** These results highlight the enhanced glucose adsorption on the heterostructure, supporting our approach of utilizing the heterojunction over pristine structures.

To elucidate the stronger glucose binding affinity to BP-gCN-O$_{14}$ compared to gCN, we calculated the projected density of states (PDOS) for these surfaces. The analysis focused on the short bonds between glucose and surface atoms (≤ 2.1 Å), which are indicative of strong interactions. **Figure S12a,b (Supporting Information)** illustrates that for gCN, one H-N bond is formed with a bond length of 1.87 Å, whereas for BP-gCN-O$_{14}$, four H-O bonds are formed with bond lengths ranging from 1.73 to 2.11 Å. Accordingly, in **Figure S12c,d (Supporting Information)**, we analyzed the PDOS of the H atoms participating in the strong binding to gCN and BP-gCN-O$_{14}$, along with the relevant surface atoms. It can be observed that there is some degree of hybridization between the N states of gCN and the H states of glucose. However, there is more a pronounced hybridization between the O states of BP-gCN-O$_{14}$ and H states of glucose, resulting in stronger glucose adsorption. Additionally, we performed Bader charge analysis to investigate the charge transfer between the adsorbed glucose molecule and the catalyst surfaces. The results show that the gCN surface slightly loses electrons (+0.02 |e|), while the BP-gCN-O$_{14}$ surface gains electrons (-0.14 |e|) upon glucose adsorption, as illustrated in the electron density difference plots in **Figure 6c,d.** Since the sensor sensitivity typically depends on the charge transfer amount after adsorption, [65] these results reveal that BP-gCN-O$_{14}$ exhibits better glucose sensing ability compared to pristine gCN. Overall, our theoretical findings provide a comprehensive explanation for the superior glucose sensing performance of oBP-gCN observed experimentally.

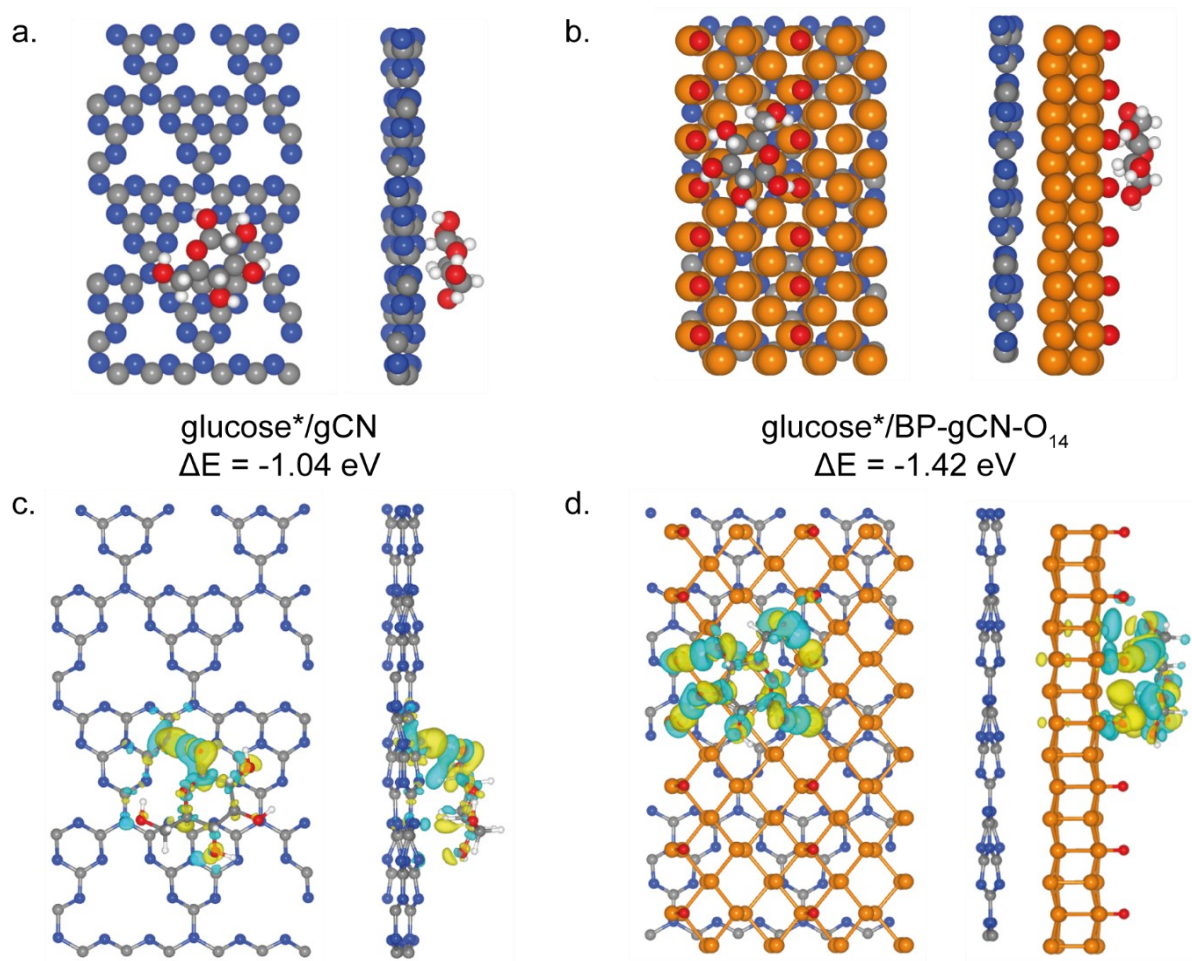

**Figure 6. Theoretical analysis of glucose adsorption on electrocatalyst surfaces.** Top and side views of glucose adsorption on **a)** wrinkled gCN and **b)** BP-gCN-$O_{14}$ with corresponding adsorption energies (ΔE). Differential charge densities of **c)** gCN and **d)** BP-gCN-$O_{14}$ after glucose adsorption. Yellow and cyan isosurfaces (±0.001 e/Bohr$^3$) represent charge accumulation and depletion, respectively. Atom color codes: blue (N), gray (C), orange (P), red (O), white (H).

## 3. Conclusion

This study presents a flexible, non-enzymatic wearable glucose sensor based on BP-gCN heterostructures, offering a robust platform for real-time, non-invasive sweat glucose monitoring. The integration of rational material design and wearable technology enabled the development of a high-performance sensing device with a sensitivity of 1.1 µA mM$^{-1}$ cm$^{-2}$ under physiological conditions. The superior glucose oxidation activity of BP-gCN heterostructures was attributed to the engineered P-N interactions, which significantly enhanced the electrochemically active surface area and reduced charge transfer resistance. Complementary DFT calculations provided further insights into optimized electronic properties of the heterostructures, demonstrating stronger glucose adsorption and enhanced charge transfer on the BP-gCN surface compared to pristine gCN. The device's applicability was also validated through on-body tests, where the conformal wearable patch enabled reliable, wireless glucose monitoring via an NFC-enabled smartphone application. The robust performance, high sensitivity, and stability under physiological conditions highlight the potential of BP-gCN heterostructures for addressing challenges in continuous glucose monitoring. This work not only demonstrates a scalable and biocompatible platform for sweat-based glucose sensing but also paves the way for the development of advanced nanomaterial-based wearable sensors. Future research can explore novel metal-free nanomaterials for higher sensitivity or extending their application to multi-analyte detection, unlocking new possibilities for next-generation wearable biosensing technologies tailored for personalized health monitoring.

## 4. Experimental Section/Methods

The experimental details are provided in the Supporting Information.

**Acknowledgments**

The authors acknowledge the use of the services and facilities of Koç University Surface Science and Technology Center (KUYTAM) and n$^2$STAR-Koç University Nanofabrication and Nanocharacterization Center for Scientific and Technological Advanced Research. Computational part of this work was performed at High Performance Grid Computing Center (Tr-Grid e-infrastructure) TUBITAK ULAKBIM. E.E.O. is supported by TUBITAK through the 2211/A program. Ö.M acknowledges the Turkish Academy of Sciences (TUBA) for the partial financial support. L.B. acknowledges the Scientific and Technological Research Council of Turkey (TUBITAK) as the project received funding from 2232 program (grant no: 118C295) and 3501 program (grant no: 120M363).

# Supporting Information

**High performance Black Phosphorus/Graphitic Carbon Nitride Heterostructure-based Wearable Sensor for Real-time Sweat Glucose Monitoring**

*Ecem Ezgi Özkahraman[a,b], Zafer Eroğlu[b], Vladimir Efremov[b], Arooba Maryyam[c], Taher Abbasiasl[d], Ritu Das[e], Hadi Mirzajani[e], Berna Akgenc Hanedar[g,h], Levent Beker[d,e]\*, Onder Metin[a,b,f]\**

[a]Department of Materials Science and Engineering, College of Engineering, Koç University, 34450 Istanbul, Turkiye

[b]Department of Chemistry, College of Sciences, Koç University, 34450 Istanbul, Turkiye

[c]Department of Electrical and Electronics Engineering, College of Engineering, Koç University, 34450 Istanbul, Turkiye

[d]Department of Biomedical Sciences and Engineering, College of Engineering, Koç University, 34450 Istanbul, Turkiye

[e]Department of Mechanical Engineering, College of Engineering, Koç University, 34450 Istanbul, Turkiye

[f]Koç University Surface Science and Technology Center (KUYTAM), 34450 İstanbul,

[g]Department of Physics, College of Sciences, Koç University, 34450 Istanbul, Turkiye

[g]Department of Physics, College of Sciences, Kirklareli University, 39100 Kirklareli, Turkiye

\*To whom should be corresponded: Prof. O. Metin, e-mail: ometin@ku.edu.tr, Prof. L. Beker, e-mail: lbeker@ku.edu.tr

## 1. Materials

Urea, red phosphorus (98.9%), tin (99.5%), tin (IV) iodide (95%), sodium hydroxide (NaOH), potassium chloride (KCl), potassium hexacyanoferrate ($K_3Fe(CN)_6$), ascorbic acid ($C_6H_8O_6$), and glycine were purchased from Sigma-Aldrich (MO, USA). Dimethylformamide (DMF) and absolute ethanol (EtOH) were obtained from Merck (Germany). Glucose and ACS grade sucrose were acquired from neoFroxx (Hesse, Germany) and Caisson Labs (USA), respectively. Nafion was sourced from Alfa Aeser (Germany). The glassy carbon electrode (GCE) and Dropsense 110 electrodes were obtained from Metrohm (USA). Phosphate-buffered saline (PBS; 1×, pH 7.2) was purchased from Thermo Fisher Scientific (MA, USA). Conductive silver and carbon pastes were obtained from DuPont (DE, USA) for lab-made electrodes and wearable devices. A single-sided flexible PCB layer (AC092500E, Pyralux) was purchased from DuPont (USA). Transparent PET films were purchased from MG Chemicals (Burlington, Canada). All chemicals were used as received, and ultra-pure water was utilized in all experimental processes.

## 2. Instruments

X-ray photoelectron spectroscopy (XPS) data were acquired using a Thermo K-alpha X-ray photoelectron spectrometer (Thermo Fisher Scientific, USA). Scanning electron microscopy (SEM) imaging, along with energy-dispersive X-ray spectroscopy (EDS), was performed on a Zeiss Ultra Plus field emission scanning electron microscope (Zeiss, Germany). Raman spectroscopy was conducted using a Renishaw Invia Raman microscope (Renishaw, UK). The crystal structure of the materials was investigated via X-ray diffraction (XRD) using a Bruker D2 Phaser X-ray diffractometer (Bruker, Germany). Ultraviolet-visible and near-infrared (UV-Vis-NIR) absorption spectra were recorded using a Shimadzu UV-3600 spectrophotometer (Shimadzu, Japan). The morphology of the synthesized materials was further examined with transmission electron microscopy (TEM) using a Hitachi HT7800 operating at 120 kV (Hitachi, Japan). The thickness of the nanosheets were characterized using a Bruker Dimension Icon atomic force microscope (AFM) (Bruker, Germany). Photoluminescence (PL) spectroscopy data were obtained using an Agilent Cary Eclipse PL (Agilent Technologies, USA). Time-resolved photoluminescence spectroscopy was performed with an Edinburgh Instruments

FLS1000 spectrometer (Edinburgh Instruments, UK). Electrochemical tests were conducted using a CHI 660E and a Gamry 1010 Potentiostats (CH Instruments, USA; Gamry Instruments, USA). Mass spectrometry was conducted using a Waters Vion IMS Q-TOF mass spectrometer (Waters Corporation, USA).

## 3. Synthesis of gCN and Exfoliation Procedure

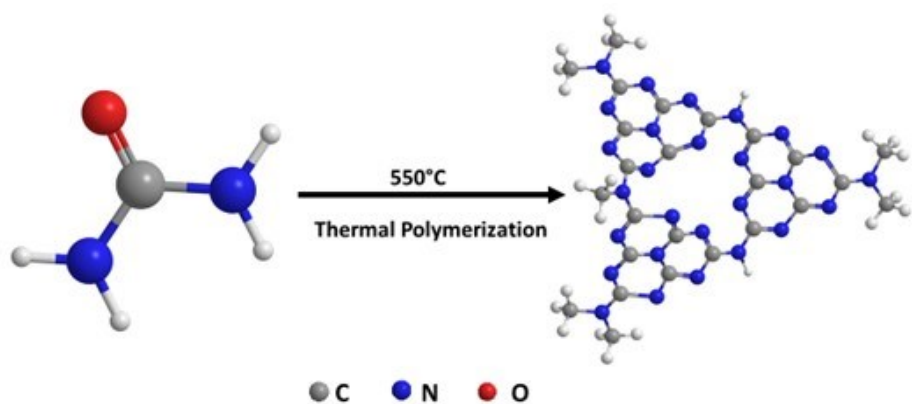

**Schematic S1.** Synthesis procedure of gCN from Urea.

gCN was synthesized via the thermal polymerization method, as shown in **Schematic S1**, using urea as the precursor. Firstly, urea powder was ground and placed in an aluminum crucible.[1] The sample was then subjected to thermal treatment in an oven with a ramping rate 5 °C/min up to 550°C and kept at this temperature for 2 hours under ambient conditions. Obtained bulk gCN powder stored for future use. Since bulk powder has limited surface area, it is necessary to exfoliate gCN to its nanosheets. To produce gCN nanosheets, exfoliation was carried out using an ultrasonic bath in a solution of 1:3 isopropyl alcohol (IPA) to deionized water (DI) for 1 hour with bulk gCN at a concentration of 0.5 mg/mL.

## 4. Synthesis of Black Phosphorus (BP) and Exfoliation Procedure

Three-dimensional (3D) layered black phosphorus (3D-BP) crystals were synthesized using the chemical vapor transport method known in the literature and previously used by our research group.[2] In a typical 3D-BP synthesis procedure, $SnI_4$ (10 mg), Sn (20 mg) and 500 mg of Red Phosphorus were transferred into a previously custom-ordered quartz ampoule and then the ampoule was sealed under vacuum. First, the temperature was raised to 620°C over 5 hours and kept at this level for 2 hours. The ampoule was then cooled to 485°C and kept at this temperature for 2 hours, then allowed to cool naturally to room temperature. Once the ampoule reached room temperature, the ampoule was broken in toluene to collect BP crystals. BP crystals were sonicated in absolute ethanol for 30 min to remove surface impurities.

Purified 3D-BP crystals were transferred to a vacuum-sealed container for later use. Then, 3D-BP, which has multiple layers of 2D nanosheets, was exfoliated by ultrasonic sound waves-assisted liquid phase exfoliation in DMF. Exfoliated BP was stored to be used in heterostructure preparation.

**5. Preparation of BP-gCN Heterostructure**

For exploring the optimal ratio between exfoliated BP to gCN, different percentages of exfoliated BP addition to gCN were prepared and characterized.[3] The formation of heterostructure was achieved by the slow addition of exfoliated BP into gCN nanosheets. Mixture of two solutions was sonicated for 3h. Final solution was centrifuged at 12000 rpm and precipitate phase collected and washed with ethanol two times by centrifugation. By the same method 5, 10 15 and 20 % (w/w %) of BP-gCN heterostructures were prepared and characterized.

**6. Functionalization of Electrodes**

Metrohm Dropsense-110 electrodes were used for electrochemical characterizations, along with our lab-made SPCE. Additionally, the effect of scan rate and EIS measurements were investigated on a glassy carbon electrode (GCE). Before functionalization, the electrode surface was washed with ethanol and DI water, then cleaned electrochemically in NaOH solution by cycling between -1 V and 1 V for 10 cycles. The synthesized BP-gCN heterojunctions were dispersed in an IPA–DI water mixture (1:3) at a concentration of 0.5 mg/mL by sonication for 1 hour to create a homogeneous ink. The dispersion was mixed with 5 wt% Nafion (1:10) to enhance binding to the electrode surface, and 10 µL of the final solution was drop-coated onto the working area of the SPCE. The coated electrodes were left to dry at room temperature for 1 hour, preparing them for subsequent glucose sensing applications

**7. Electrochemical Characterizations**

The electrochemical experiments were performed by CHI 660E electrochemical workstation and Gamry 1010 potentiostats. SPCE was used for the CHI experiments by 3 electrode setup: carbon as working and counter electrode and Ag as reference electrode. Working electrode area was functionalized for the glucose sensing. DPV measurements were done between -0.2 V and 0.4 V. CV measurements done between -1V and +1V in PBS with scan rate of 0.1 V/s. ECSA was calculated using the Randles-Sevcik equation based on CV at different scan rates. The peak current ($Ip_a$) is measured in amperes (A), $C_0$ represents the concentration of the electroactive species ($5 \times 10^{-6}$ M), n is the number of electrons involved in the redox process (1), D is the diffusion coefficient of the redox species ($6.7 \times 10^{-6}$ cm²/s), v is the scan rate (V/s), and A refers to the ECSA in cm². By substituting these values into the Randles-Sevcik equation (Equation (1)), ECSA for BP-gCN and pristine gCN was calculated.

Randles-Sevcik Equation:
$$Ip = (2.69 \times 10^5) n^{3/2} A D^{1/2} C_0 v^{1/2} \qquad \text{Eq (1)}$$

## 8. Computational Methods

Density functional theory (DFT) calculations were carried out in VASP[4] using projector augmented wave (PAW) pseudopotentials[5] and the PBE functional[6]. The energy and force convergence thresholds were set to $10^{-5}$ eV and 0.03 eV/Å, respectively, while the energy cutoff was set to 500 eV. A $2 \times 1 \times 1$ Monkhorst–Pack k-point mesh and a vacuum distance of 25 Å along the z-direction were used for all the calculations.[7] The glucose adsorption energy was calculated as follows: $\Delta E = E_{surf+glucose*} - E_{surf} - E_{glucose}$, where $E_{surf+glucose*}$, $E_{surf}$, and $E_{glucose}$ represent the total energies of the surface with adsorbed glucose molecule, the surface itself, and the glucose molecule, respectively. Bader charge analysis was conducted to explore the charge distribution between studied surfaces and adsorbed glucose molecule.[8] VESTA was used for visualization of atomic structures.[9]

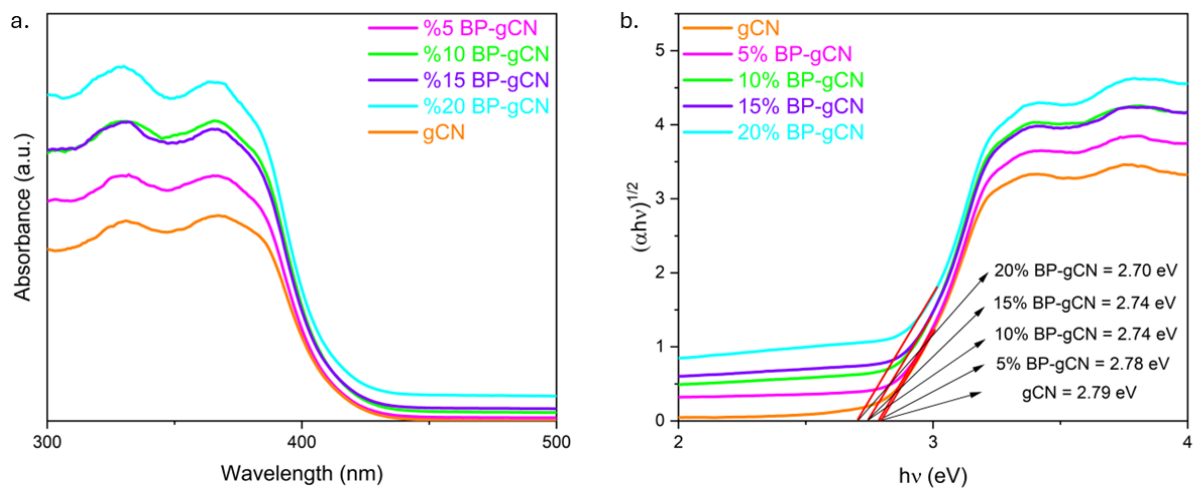

**Figure S1. Photophysical Properties of Pristine gCN and BP-gCN Heterostructures.**
**a)** UV-VIS absorbance spectrum of gCN and various wt% of BP in the BP-gCN heterostructure.
**b)** Tauc's plot derived from the absorbance data, showing the calculated bandgaps for the different compositions.

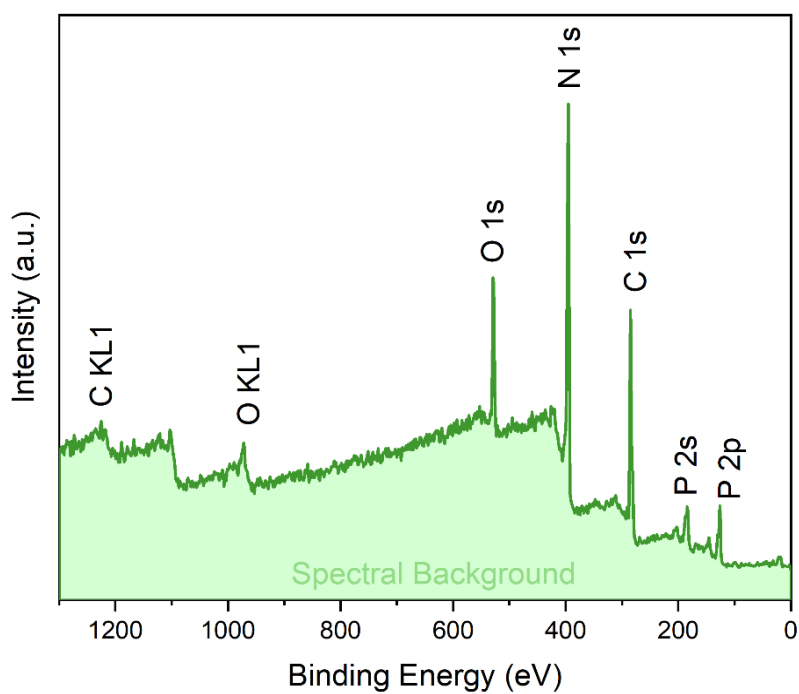

**Figure S2**. XPS survey of BP-gCN heterostructure depicting the presence of P 2p, O 1s, C 1s, N 1s species.

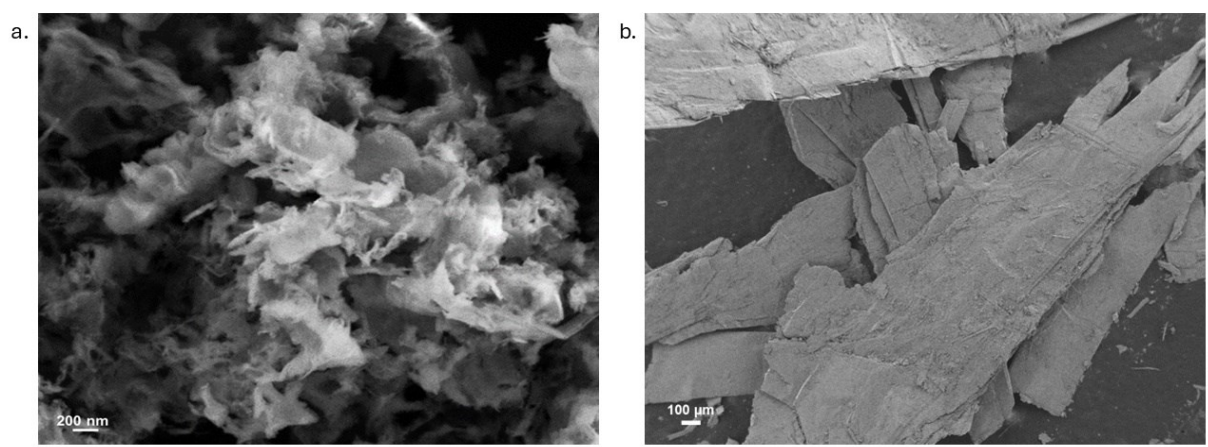

**Figure S3. SEM images of the bulk and layered pristine structures.**
 **a)** gCN powder and **b)** BP powder, images showcasing their layered structure at various magnifications.

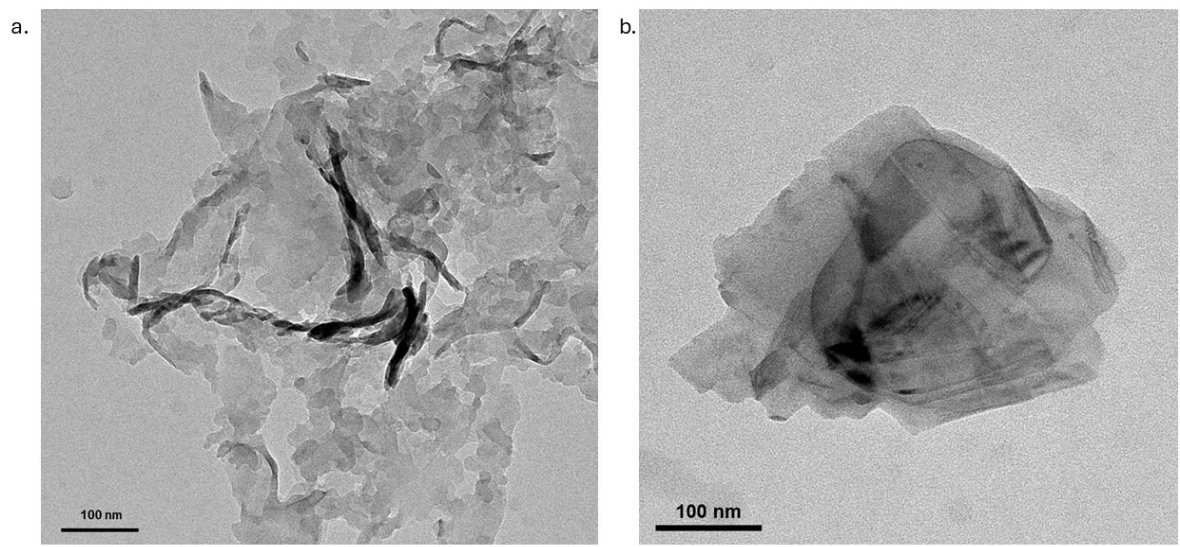

**Figure S4. TEM images of pristine structures.**

Exfoliated nanosheets of **a)** gCN and **b)** BP at 100 nm scale.

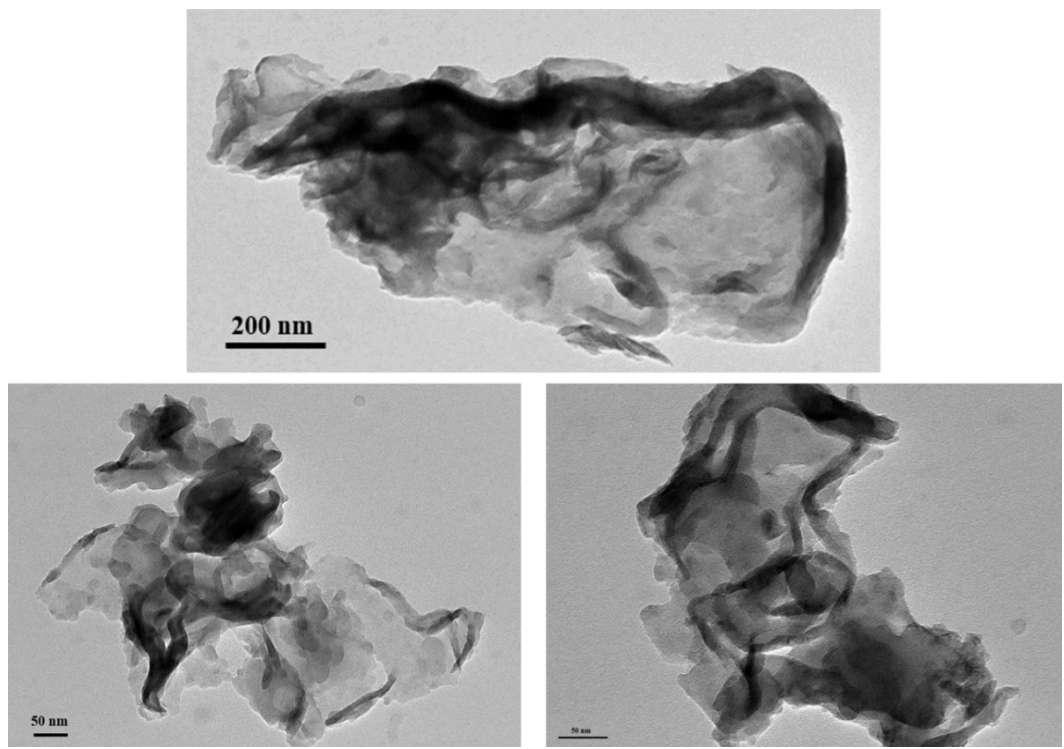

**Figure S5.** TEM images of BP-gCN heterostructure nanosheets at 200 nm and 50 nm scales.

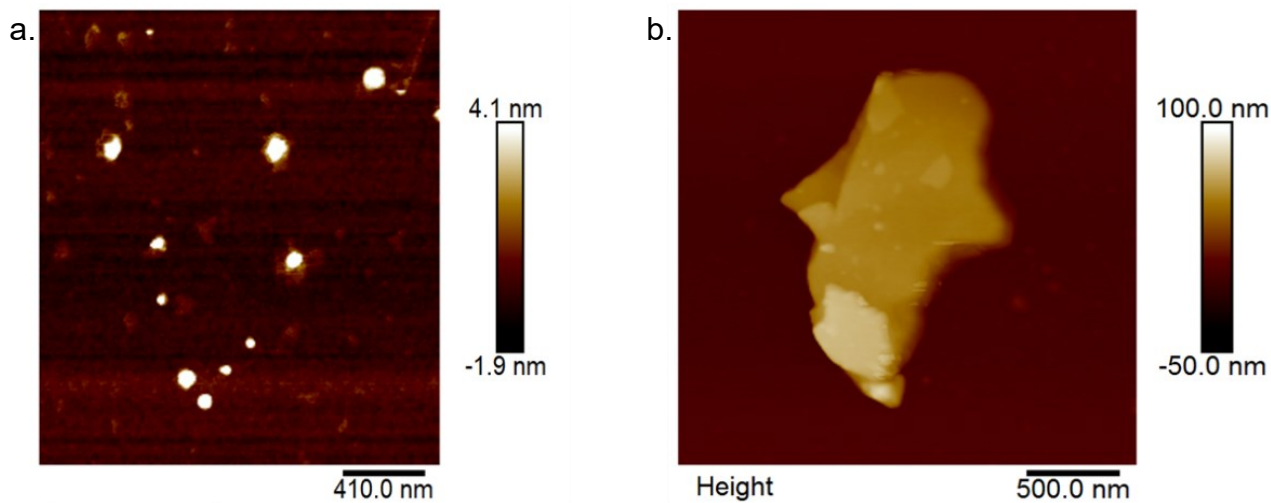

**Figure S6. AFM Images and Height Profiles.**

**a)** gCN and **b)** BP nanosheets illustrating their surface topography and layer thicknesses at the nanoscale.

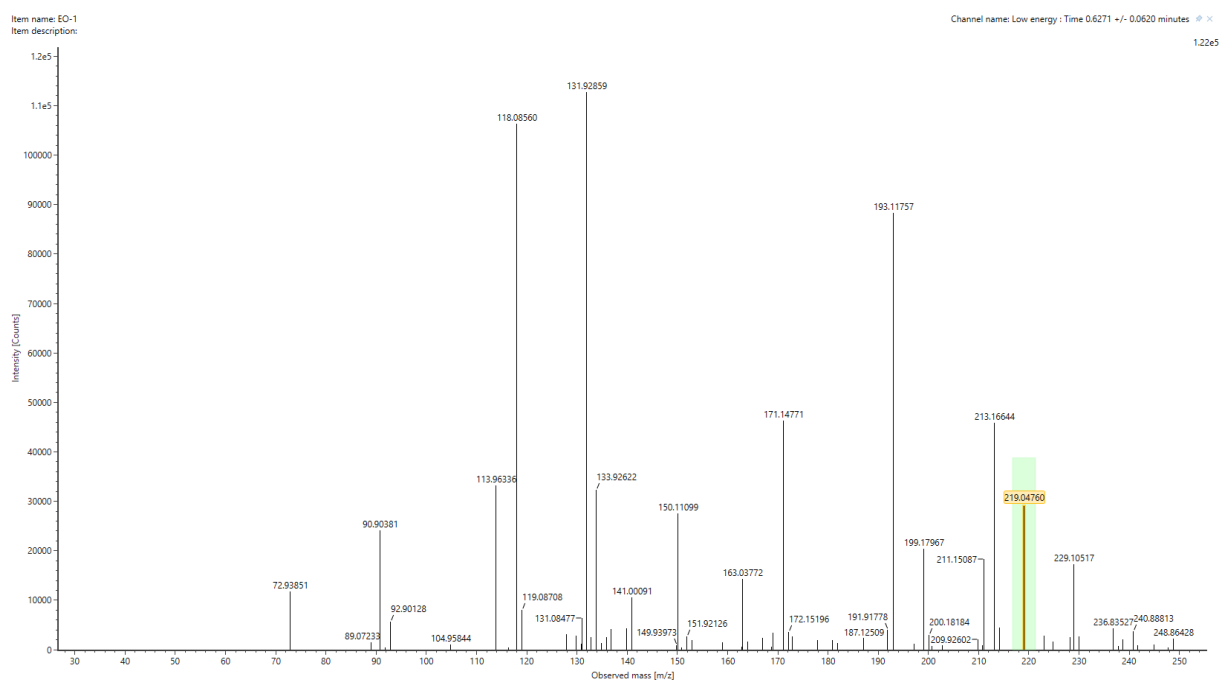

**Figure S7.** qTOF-MS Spectrum showing gluconic acid peak at 219.04 m/z.

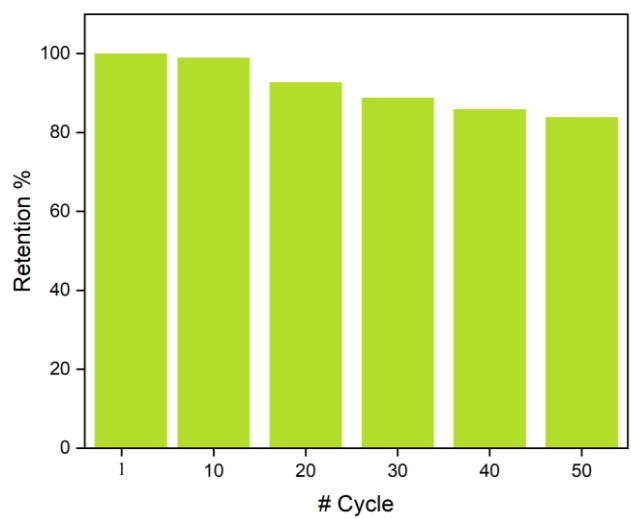

**Figure S8.** Stability of oBP-gCN for over 50 cycles.

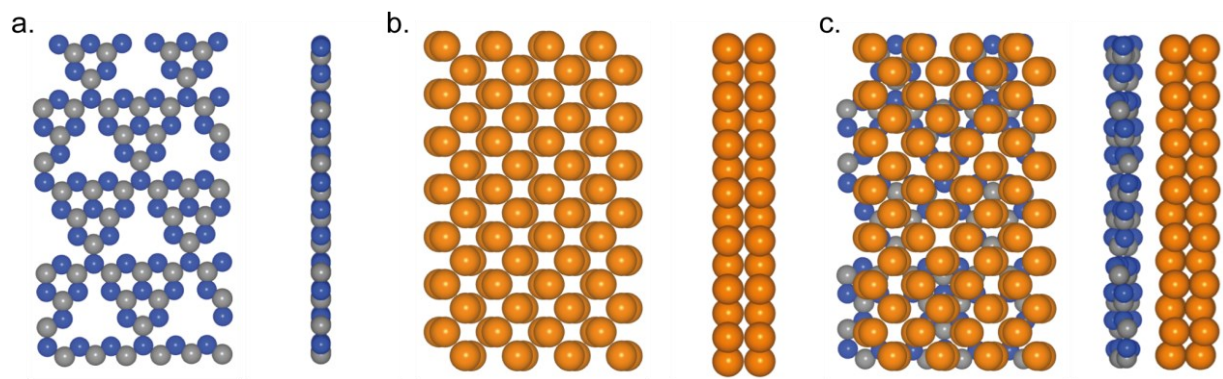

**Figure S9. Top and side views of the optimized structures.**

**a)** Planar pristine gCN, **b)** planar pristine BP, and **c)** BP-gCN heterostructure. Atom color codes: Blue (N), Grey (C), Orange (P).

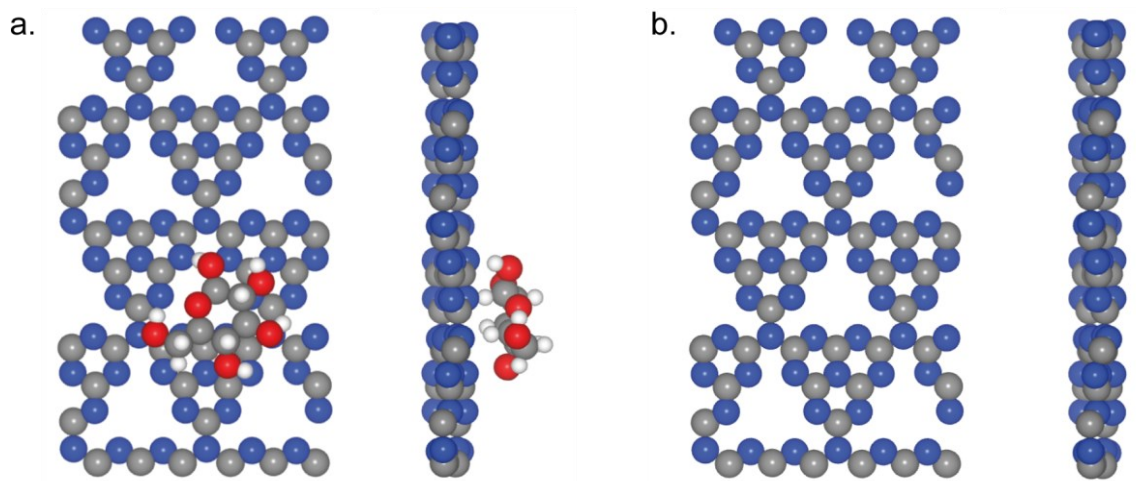

**Figure S10. Wrinkled gCN structure formation upon glucose adsorption.**

Top and side views of **a)** glucose adsorption on wrinkled gCN and **b)** wrinkled gCN after the removal of adsorbed glucose. Atom color codes: Blue (N), Grey (C), Red (O), White (H).

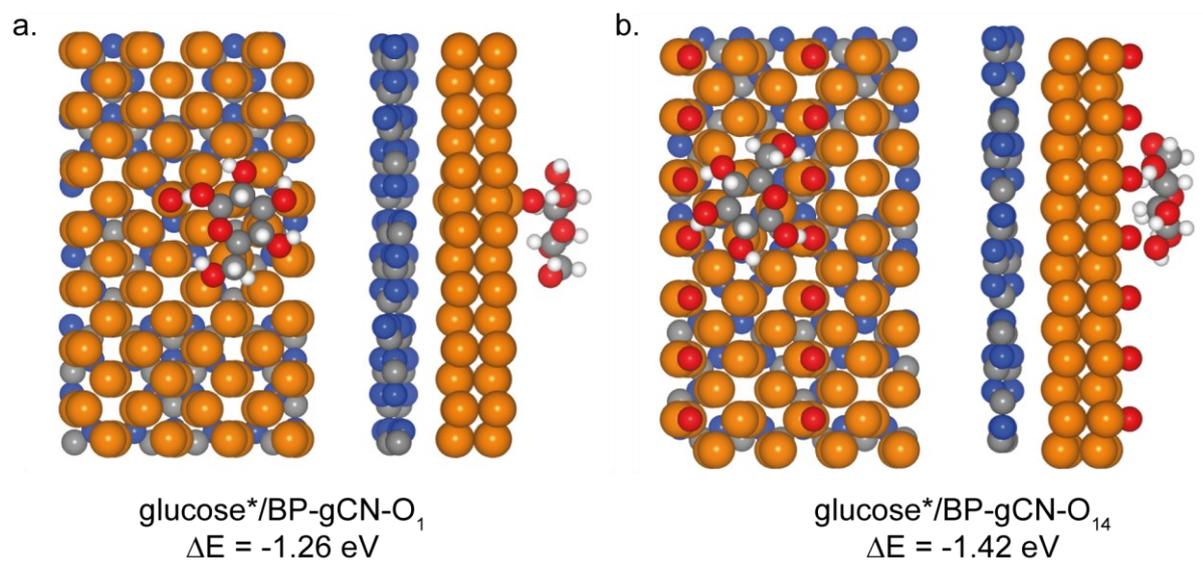

**Figure S11. Model structures used in the adsorption energy calculations.**
Top and side views of glucose adsorption on **a)** BP-gCN-O$_1$ and **b)** BP-gCN-O$_{14}$ with corresponding adsorption energies (ΔE). Atom color codes: Blue (N), Grey (C), Orange (P), Red (O), White (H).

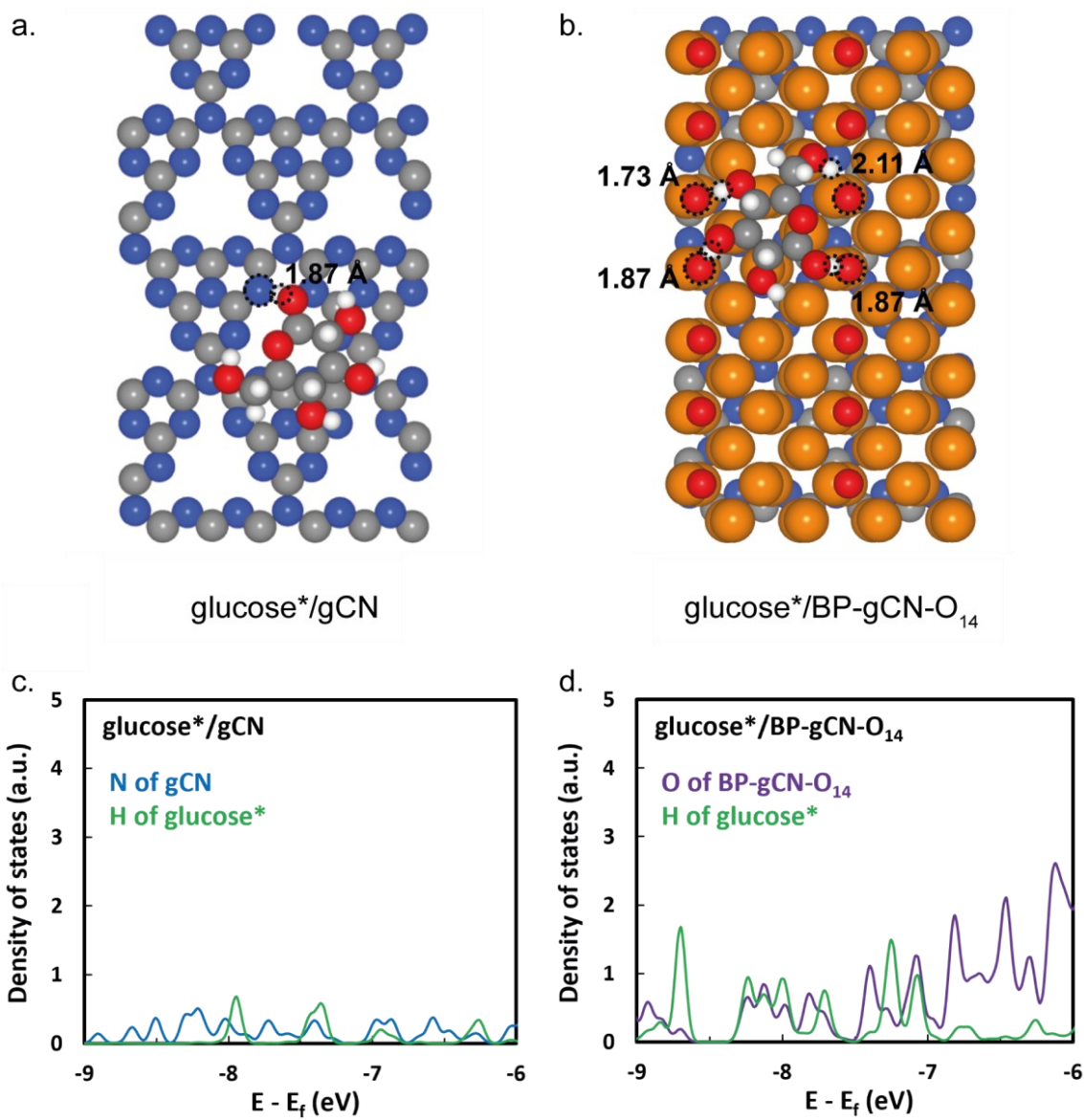

**Figure S12. Projected density of states (PDOS) analysis.**

Bond lengths between glucose and surface atoms (highlighted by circles) for **a)** gCN and **b)** BP-gCN-O$_{14}$. Only the bond lengths ≤ 2.1 Å are shown. PDOS analysis of **c)** gCN for the atoms highlighted in (a) and **d)** PDOS analysis of BP-gCN-O$_{14}$ for the atoms highlighted in (b). The Fermi level ($E_F$) is set to zero. Atom color codes: Blue (N), Gray (C), Orange (P), Red (O), White (H).